\documentclass[fleqn,usenatbib]{mnras}

\usepackage{newtxtext,newtxmath}

\usepackage[T1]{fontenc}

\DeclareRobustCommand{\VAN}[3]{#2}
\let\VANthebibliography\thebibliography
\def\thebibliography{\DeclareRobustCommand{\VAN}[3]{##3}\VANthebibliography}

\newcommand{\soutPC}{\bgroup\markoverwith{\textcolor{cyan}{\rule[0.5ex]{2pt}{1pt}}}\ULon}

\newcommand{\soutdif}{\bgroup\markoverwith{\textcolor{magenta}{\rule[0.5ex]{2pt}{1pt}}}\ULon}

\usepackage{graphicx}	
\usepackage{amsmath}	
\usepackage{caption}
\usepackage{xcolor}
\usepackage{subcaption}
\usepackage[normalem]{ulem} 

\defcitealias{Weller:2022aa_TNG50}{W22}
\defcitealias{Weller:2022aa_Astrid}{W23}
\defcitealias{Ricarte:2021aa}{R21}
\defcitealias{Donkelaar:2022ab}{Paper~I}
\defcitealias{Donkelaar:2023aa}{Paper~II}

\title[Wandering IMBHs in cosmological simulations]{Wandering intermediate-mass black holes in Milky Way-mass galaxies in cosmological simulations: myth or reality?}
\author[F. van Donkelaar et al.] 
{Floor van Donkelaar,\thanks{floor.vandonkelaar@uzh.ch} Lucio Mayer, Pedro R. Capelo and Tomas Tamfal\\Department of Astrophysics, University of Zurich, Winterthurerstrasse 190, CH-8057 Z\"urich, Switzerland}

\date{Accepted XXX. Received YYY; in original form ZZZ}

\pubyear{2025}

\begin{document}
\label{firstpage}
\pagerange{\pageref{firstpage}--\pageref{lastpage}}
\maketitle

\begin{abstract}
In this work, we address the question\textit{``can we use the current cosmological simulations to identify intermediate-mass black holes (IMBHs) and quantify a putative population of wandering IMBHs?''} and demonstrate that caution is necessary when drawing conclusions about wandering IMBHs in the Milky Way based on cosmological simulations, due to their relatively low resolution. We compare wandering-IMBH counts in different simulations with different sub-grid methods and post-processing recipes, the ultimate goal being to aid future wandering-IMBH detection efforts. In particular, we examine simulations in which IMBHs are identified as BH seeds forming at high redshift and those in which they are identified using star clusters as proxies, which implicitly appeals to a stellar dynamical formation channel. In addition, we employ the extremely high-resolution cosmological hydrodynamical ``zoom-in'' simulation of a Milky Way-sized galaxy, GigaEris, with the star cluster proxies method to identify IMBHs.  Wandering IMBHs are defined using two methods: those within but originating beyond the virial radius, and ``non-central'' BHs. We find consistent counts of wandering high-redshift IMBHs across most of the different cosmological simulations employed so far in the literature, despite the different identification approaches, resulting in 5 to 18 wandering IMBHs per Milky Way-sized galaxy at $z \geq 3$. Nevertheless, we argue this is only coincidental, as a significant discrepancy arises when examining the formation sites and the mass ranges of the wandering IMBHs.  This raises questions about the extent to which current cosmological simulations can guide observational searches for wandering IMBHs.
\end{abstract} 

\begin{keywords}
black hole physics – methods: numerical –  software: simulations
\end{keywords}



\section{Introduction}

Black holes (BHs) are commonly detected throughout the Universe and cover a large mass range. The extensively detected BHs can be mainly divided into two populations at the extremes of their mass range: stellar-mass BHs and supermassive BHs (SMBHs). Stellar-mass BHs typically have masses $M_{\bullet} \lesssim 10^3$ M${_{\sun}}$ and are found throughout galaxies; for example, the Milky Way (MW) is expected to host $\sim$$10^8$ stellar-mass BHs \citep[e.g.][]{Pods:2003aa, Farr:2011aa, Elbert:2018aa}. SMBHs, on the other hand, have masses $M_{\bullet} \gtrsim 10^6$ M${_{\sun}}$ and are usually found at the centre of massive galaxies. SMBHs are tiny compared to galactic scales but can manifest themselves as powerful active galactic nuclei \citep[AGN; e.g.][]{Kormendy:2013aa} fed by gas accretion, which can make them outshine their host galaxy. The MW has a central SMBH with a mass of $\sim$$4 \times 10^6$~M${_{\sun}}$ \citep[e.g.][]{Ghez:2008aa, Genzel:2010aa}.

Intermediate-mass BHs (IMBHs) bridge, as the name suggests, the gap between stellar-mass BHs and SMBHs and fall within the mass range $10^3$ M${_{\sun}} \lesssim M_{\bullet} \lesssim 10^6$~M${_{\sun}}$. Observationally, IMBHs have proven challenging to detect, unlike their counterparts, the stellar-mass BHs and SMBHs. However, in recent years there has been growing observational evidence for the existence of IMBHs with masses bordering close to the lower and upper-mass limits. Several observations have indicated the presence of IMBHs in galactic nuclei. For instance, \citet{Kunth:1987aa} pointed out that an AGN in the dwarf galaxy POX 52 might be powered by a $\sim$$10^5$~M${_{\sun}}$ BH \citep[see][]{Barth:2004aa, Rizzuto:2023aa}. Similarly, a $10^4$--$10^5$~M${_{\sun}}$ BH could be responsible for the activity detected in the nuclear star cluster of the dwarf  galaxy NGC~4395 \citep{Filippenko:1989aa, Shih:2003aa,  Peterson:2005aa, denbrock:2015aa, Rizzuto:2023aa}. In recent times, an increasing number of candidate IMBHs have been (indirectly) detected within dwarf galaxies through various techniques, such as the analysis of broad and narrow emission lines, infrared emission, and X-ray observations \citep[see, e.g.][]{Dewangan:2008aa, Miniutt:2009aa, Reines:2011aa, Kamizasa:2012aa, Baldassare_et_al_2015, Sartori:2015aa, Mezcua:2017aa, Reines:2020aa, Greene:2020aa, Smith:2023aa, Barrows:2024aa, Chrisholm:2024aa}. 

Another place where we could possibly find IMBHs is within our own Galactic Centre \citep[e.g.][]{Greene:2020aa}. The MW could host a population of ``leftover'' IMBHs from past accretion events of dwarf galaxies \citep[see, e.g.][]{Rashkov:2014aa}. In the present, IMBHs have been linked to several observational phenomena associated with the Galactic Centre, although none of the evidence can be regarded as definitive \citep[][]{Greene:2020aa}. Additionally, studies in the last few years have revealed that a considerable amount of BHs in galaxies experience ineffective dynamical friction (DF) and do not drift to the centre of the galaxy halo \citep[e.g.][]{Bellovary:2010aa, Capelo_et_al_2015, Volonteri:2016aa, Tamburello_et_al_2017, Tamfal_et_al_2018, Tremmel:2018aa, Tremmel:2018ab, Pfister:2019aa, Bortolas:2020aa, Bellovary:2021aa, Chen:2022aa}. This could lead to a population of ``wandering'' IMBHs (see, e.g. \citealt{Ricarte:2021aa}, hereafter \citetalias{Ricarte:2021aa}; \citealt{Ricarte:2021ab}; \citealt{Weller:2022aa_TNG50}, hereafter \citetalias{Weller:2022aa_TNG50}; \citealt{Weller:2022aa_Astrid}, hereafter \citetalias{Weller:2022aa_Astrid}). Furthermore, galaxy-galaxy mergers could likewise increase the potential of IMBHs wandering in galaxies like the MW, which are estimated to have experienced 15~$\pm$ 3 mergers, and 9 $\pm$ 2 of these taking place at $z > 2$, with dwarf galaxies with stellar masses $\geq 4.5 \times 10^6$~M${_{\sun}}$ \citep[][]{Kruijssen:2019aa, Kruijssen:2020aa}.

IMBHs possibly wandering within the MW might have originated from two sources: (i) formation within the galaxy (in situ) and (ii) formation outside the galaxy (ex-situ). In-situ formation pathways involve the direct collapse of high-mass quasi-stars \citep[e.g.][]{Volonteri:2010ab, Schleicher:2013aa}, super-Eddington accretion onto stellar-mass BHs \citep[e.g.][]{Ryu:2016aa,Sassano_et_al_2023}, runaway mergers in dense globular stellar clusters \citep[e.g.][]{Portiegies:2004aa, Portiegies:1999, Devecchi:2009aa, Mapelli:2016aa, shi:2021aa,Gonzalez:2021aa}, and supra-exponential accretion onto seed BHs in the early Universe \citep[e.g.][]{Alexander:2014aa, Nata:2021aa}. The ex-situ pathway produces these roaming BHs, such as through events like the tidal disruption of satellite or dwarf galaxies merging into larger halos \citep[e.g.][]{Volonteri:2003aa, Oleary:2009aa, Greene:2021aa}. However, there is expected to be some overlap between the pathways. Dense stellar clusters could also form beyond the virial radius of a MW-like galaxy, for example in a dwarf galaxy or completely outside dark matter (DM) halos \citep[see, e.g.][]{Madau:2020aa, Lake:2023aa, vandonkelaar:2023aa}.

In order to inform future observational efforts to detect IMBHs wandering in the MW, it is important to understand their dynamics and kinematics. The accretion rates of wandering IMBHs are expected to be rather negligible due to their relatively low mass, the low gas density in the galaxy halos they wandered into, and the high relative velocities of the IMBHs themselves with respect to the surrounding gas \citep[see, e.g.][]{Bellovary:2021aa}, as the relative velocity is much larger for an IMBH drifting through its host galaxy compared to one residing stationary at the centre. As a consequence, the luminosity of wandering IMBHs could be so low that they will either escape detectability or be easily confused with another kind of high-energy source, such as an X-ray binary \citep[e.g.][]{Bellovary:2021aa}. Furthermore, it remains uncertain how many of the IMBHs formed in the MW will still be wandering at $z=0$ as opposed to how many will have accreted to the central supermassive BH, contributing to its growth

Nevertheless, \citet{Seepaul:2022aa} recently investigated how $10^5$~M$_{\sun}$ IMBHs behave in different MW environments. Their findings suggest that roughly $\sim$27~per cent of these wandering IMBHs could be detected using Chandra in X-rays \citep{Weisskopf:2000aa}, $\sim$37~per cent with the Roman Space Telescope in near-infrared \citep{Wang:2022aa}, $\sim$49~per cent with CMB-S4 in sub-mm \citep{Abazajian:2016aa}, and $\sim$57~per cent using ngVLA in radio wavelengths \citep{Murphy:2018aa}. They conclude that the brightest fluxes are emitted by IMBHs passing through molecular clouds or cold neutral medium, where they are always detectable.  The likelihood of detection is set to increase significantly due to the introduction of new state-of-the-art instruments such as the Advanced X-Ray Imaging Satellite \citep[AXIS; see, e.g. ][]{Pacucci:2023aa}. Additionally, future instruments like the Laser Interferometer Space Antenna \citep[LISA;][]{Barack_et_al_2019,Amaro-Seoane_et_al_2023,Colpi_et_al_2024} and the Einstein Telescope \citep{Punturo:2010aa} offer prospects for detecting gravitational waves (GW) from tidal disruption events involving wandering IMBHs \citep[e.g.][]{Toscani:2023aa}. While the Einstein Telescope targets sources with total masses in the range of hundreds or a few thousand M${_{\sun}}$ \citep{Gair:2011aa}, LISA aims to detect GWs from systems with BH masses spanning from ten thousand to ten million M${_{\sun}}$ \citep{Gair:2011ab}. Together, these telescopes cover the expected mass range of IMBHs wandering throughout the MW.

To guide these future detections, numerous studies have delved into investigating the characteristics of the as-yet-undetected population of wandering BHs through simulation techniques (e.g. \citealt{Bellovary:2010aa, Gonzalez:2018aa, Tremmel:2018aa, Greene:2021aa}; \citetalias{Ricarte:2021aa}; \citealt{Ricarte:2021ab, Seepaul:2022aa}; \citetalias{Weller:2022aa_TNG50}; \citetalias{Weller:2022aa_Astrid}; \citealt{Dimatteo:2022aa}). Due to the low mass of IMBHs and the resolution limit of simulations, these studies predominantly adopt one of two distinct sub-grid methodologies to simulate and model BHs. The first method involves the ``seeding of BHs'', entailing the initial introduction of BHs with pre-defined properties, such as mass and position, during the simulation. The second method,``stellar proxies'', employs stellar clusters within specific mass ranges to act as substitutes of IMBHs. In particular, \citetalias{Weller:2022aa_TNG50}, who use the second method, suggest that thousands of wandering BHs might be found inside a galactic halo similar to that of the MW. Nevertheless, one first needs to obtain an understanding of the influence of the resolution and the method of modelling the BHs on the number of identified wandering IMBHs, before one can make claims on the possible amount of wandering IMBHs in the MW. 

In this work, we try to shed light on the question \textit{``can we use current cosmological simulations to identify IMBHs and characterize their properties?''}. We do this by comparing the data of the identified IMBHs within the following cosmological simulations:  TNG50 \citep[][ \citetalias{Weller:2022aa_TNG50}]{Pillepich:2019aa, Nelson:2019aa}, \textsc{ASTRID} \citep[][\citetalias{Weller:2022aa_Astrid}]{bird:2022aa, Ni:2022, Chen:2022ab}, Romulus \footnote{In the Romulus simulation, the wandering BHs have a minimum mass of $10^{6}$~M$_{\sun}$, which is at the upper limit for being classified as an IMBH. At the redshift used in our work ($z = 4.4$), these BHs are still very close to this minimum mass. Therefore, we believe it is appropriate to include them in our analysis.} \citep[][\citetalias{Ricarte:2021aa}]{Tremmel:2017aa}, and the extremely high-resolution $N$-body, hydrodynamical, cosmological ``zoom-in'' GigaEris simulation  \citep{Tamfal:2022aa}. The advantage of GigaEris, and the reason why we employ it in this study, is that its baryonic mass resolution lies at the lower end of the putative mass distribution of IMBHs, whereas all the other simulations, being ``uniform-volume", are at best resolving the upper end of such mass distribution (see next section). The layout of this paper is the following: Section~\ref{sec:method} briefly summarizes the simulation setup of GigaEris and discusses the selection criteria used for identifying IMBHs within the different simulations. In Section~\ref{sec:results}, we compare the number of  detected IMBHs in the different simulations and study their properties, including their birth location. We discuss our results and conclude in Section~\ref{sec:disc}. 

\section{Methods}\label{sec:method}

\subsection{Simulation code and initial conditions}\label{sec:sim}

To analyze wandering IMBHs, we utilize the $N$-body, hydrodynamical, cosmological ``zoom-in'' simulation of a MW-like galaxy known as GigaEris \citep[][]{Tamfal:2022aa}. This simulation was conducted using the $N$-body smoothed-particle hydrodynamics code \textsc{ChaNGa} \citep[][]{Jetley:2008aa, Jetley:2010aa, Menon:2015aa}. A brief overview of the numerical approach is presented below, while a more comprehensive description can be found in \citet{Tamfal:2022aa}. We point out that GigaEris does not contain a sub-grid model for the generation and growth of BHs. As a result, we use star clusters as proxies for BHs, following a similar approach to that in \citetalias{Weller:2022aa_TNG50} (see Section~\ref{giga_selection}). This approach enables a direct comparison between the simulations.

The simulation follows a Galactic-scale halo identified in a low-resolution, DM-only simulation at $z = 0$ in a periodic cube of side 90~cMpc \citep[see][]{Tamfal:2022aa}. The halo was deliberately chosen to match the mass of that of the MW at $z = 0$ and to have a relatively calm late merging history. This approach mirrors the selection process used for the galaxy halo in the original Eris suite detailed by \citet{Guedes:2011aa}. Nevertheless, it is important to note that the expected total mass of the selected halo in the GigaEris simulation at $z = 0$ is $1.4 \times 10^{12}$~M$_{\sun}$, which is slightly higher than the most recent estimates of the mass of the MW halo \citep[see, e.g.][]{Wang:2020aa, Bobylev:2023aa, Jiao:2023aa}. Following, the selected halo was re-simulated at several orders of magnitude higher resolution than the DM-only simulation, adding gas particles as well as the necessary short-wavelength modes to the simulation.  

In the code, a star particle represents an entire stellar population with its own \citet{Kroupa:2001aa} stellar initial mass function. Star particles are created stochastically with an initial mass of $m_{\star} = 1026$~M$_{\sun}$, using a simple gas density and temperature threshold criterion \citep[][]{Stinson:2006aa}, with $n_{\rm SF} >  100$~H~atoms~cm$^{-3}$ and $T < T_{\rm SF} = 3 \times 10^4$~K, and with a star formation (SF) rate given by

\begin{equation}
    \frac{\rm{d} \rho_{\star}}{\rm{d}t} = \epsilon_{\rm SF} \frac{\rho_{\rm gas}}{t_{\rm dyn}},
\end{equation}

\noindent where $\rho_{\star}$ represents the stellar density, $\rho_{\rm gas}$ denotes the gas density, $t_{\rm dyn}$ stands for the local dynamical time, and $\epsilon_{\rm SF}$ indicates the SF efficiency, which is fixed at $0.1$. The code solves for the non-equilibrium abundances and cooling of H and He  species, assuming self-shielding \citep[see][]{Pontzen:2008aa} and a redshift-dependent ultraviolet radiation background \citep{Haardt:2012aa}. Additionally, cooling from the fine structure lines of metals is computed in photoionization equilibrium under the same radiation background (with no self-shielding assumed, refer to \citealt{Capelo:2018aa} for discussion), utilizing tabulated rates from Cloudy \citep{Ferland:2010aa, Ferland:2013aa} and following the methodology outlined in \citet{Shen:2010aa, Shen:2013aa}. Cooling for all species is modelled down to a temperature of 10~K.

Feedback from Type Ia supernovae (SNae Ia) is implemented by injecting energy along with a fixed mass and metallicity amount, independent of the progenitor mass, into the surroundings \citep[see][]{Thielemann:1986, Stinson:2006aa}. On the other hand, feedback from Type II supernovae (SNae II) is incorporated following the delayed-cooling recipe proposed by \citet{Stinson:2006aa}, each injecting metals and $E_{\rm SN} =10^{51}$ erg per event into the interstellar medium as thermal energy, according to the ‘blastwave model’ of \citet{Stinson:2006aa}.  For each SNa II event, a given amount of oxygen and iron mass, dependent on the stellar mass, is injected into the surrounding gas \citep{Woosley:1995aa, raiteri:1996aa}. The stars with  masses between 8 and 40~M$_{\sun}$ will explode as SNae II, whereas stars with a lower stellar mass do not explode as SNae but release part of their mass as stellar winds, with the returned gas having the same metallicity of the low-mass stars.

The simulation reached $z = 4.4$, with the following particle numbers and masses in the entire simulation box at the final snapshot: $n_{\rm DM} = 5.7 \times 10^8$ for DM particles with mass $m_{\rm DM} = 5493$~M$_{\sun}$, $n_{\rm gas} = 5.2 \times 10^8$ for gas particles with mean mass $m_{\text{gas}} = 1099$~M$_{\sun}$, and $n_{\star} = 4.4 \times 10^7$ for stellar particles with mean mass $m_{\star} = 798$~M$_{\sun}$. The initial conditions were generated using the \textsc{MUSIC} code \citep[][]{Hahn:2011aa}, with 14 levels of refinement and the cosmological parameters $\Omega_{\rm m} = 0.3089$, $\Omega_{\rm b} = 0.0486$, $\Omega_{\Lambda} = 0.6911$, $\sigma_8 = 0.8159$, $n_{\rm s} = 0.9667$, and $H_0 = 67.74$~km~s$^{-1}$~Mpc$^{-1}$ \citep[see][]{Planck:2016aa}. The gravitational softening of all particles is set to a constant in physical coordinates ($\epsilon_{\rm c} = 0.043$~kpc) for redshifts smaller than $z = 10$, and otherwise evolves as $\epsilon = 11\epsilon_{\rm c}/(1+z)$.

\subsection{Selection criteria for identified intermediate-mass black holes}\label{sec:criteriamethod}

To study the identification of IMBHs in cosmological simulations, we first need to understand the differences between various simulation methods. In this work, we compare our findings with those of \citetalias{Weller:2022aa_TNG50}, \citetalias{Weller:2022aa_Astrid}, and \citetalias{Ricarte:2021aa}, and clarify the differences present in these studies: two distinct methodologies are employed to simulate and model BHs. The ASTRID (\citetalias{Weller:2022aa_Astrid}) and Romulus (\citetalias{Ricarte:2021aa}) simulations involve seeding BHs directly as part of their process. This ``seeding of BHs'' refers to the initial placement of BHs with predefined properties, such as mass and position, during the simulation. Conversely, the TNG50 (\citetalias{Weller:2022aa_TNG50}) and GigaEris works utilize stellar clusters falling within specific mass ranges to serve as proxies for IMBHs.

It is also essential to acknowledge the use of different terminologies for the BHs within these studies. Specifically, the term ``wandering IMBH'' lacks a clearly established or precise definition. In this work, we use two different definitions for wandering IMBHs (both ascertained at the final snapshot of our simulation) to accurately compare between methods. First, we define wandering IMBHs as those BHs situated within a galaxy's virial radius but originating beyond the virial radius of said galaxy,\footnote{For this latter check, we will be using the final virial radius at $z=4.4$ (see Section~\ref{giga_selection}).} similar to the definition used by \citetalias{Weller:2022aa_TNG50} \citep[but also, e.g.][]{Bellovary:2010aa, Gua:2020aa, Izquierdo:2020aa, Chitan:2022aa, Mahler:2023aa}. For comparisons to the BH-seeding method, we define as wandering IMBHs the so-called ``non-central'' BHs, i.e. those BHs that are beyond a minimum given distance from the halo's centre, as done in \citetalias{Weller:2022aa_Astrid} and \citetalias{Ricarte:2021aa}, and use the threshold of 0.7~kpc \citepalias[as in][]{Ricarte:2021aa}.

BH seeding involves introducing BHs into simulations when an environment satisfies all the predetermined criteria and characteristics. In \citetalias{Ricarte:2021aa}, the BHs are seeded according to nearby gas properties, rather than being associated with particular host halos or galaxies. This method does not guarantee that every halo hosts a BH and allows multiple BHs to form in one halo. Gas particles that have been selected for SF and additionally meet the specific criteria of having a very low mass fraction of metals ($Z < 3 \times 10^{-4}$), a high density (at least 15 times their SF density threshold of 0.2~H~atoms~cm$^{-3}$), and warm temperature (between 9500 and 10000~K, their temperature threshold for SF) transform into BHs (with a fixed seed mass of $10^6$~M$_{\sun}$). This seeding mirrors the direct-collapse BH scenario, in which extreme conditions lead to a large \citet{Jeans_1902} mass collapsing into a BH seed \citep[e.g.][]{Oh:2002aa, Bromm:2003aa, Lodato:2006aa, Begelman:2006aa}. The initial BH mass is set at $10^6$~M${_{\sun}}$ to ensure that it remains more massive than other particles in the simulation, preventing scattering events caused by interactions with smaller particles. In this work, we use the radial cut used in \citetalias{Ricarte:2021aa} for one of our sets,  meaning that all the BHs beyond 0.7 kpc from the halo centre are seen as wandering BHs. 

On the other hand, \citetalias{Weller:2022aa_Astrid} used a halo-based seeding model, in which the BHs are seeded in halos with $M_{\rm halo} > 7.38 \times 10^9$~M${_{\sun}}$ and $M_{\rm star}> 2.95 \times 10^6$~M${_{\sun}}$. The value of $M_{\rm star}$ is a conservative choice to make sure that BHs are seeded in halos with enough cold dense gas to form stars, and that there are at least some collisionless star particles in the BH neighbourhood to act as sources of DF. The seed masses in \citetalias{Weller:2022aa_Astrid} are stochastically drawn from a power-law probability distribution, with a mass between $4.43 \times 10^4$ and $4.43 \times 10^5$~M${_{\sun}}$. In both works, the dynamics of the BHs are modelled with the same sub-grid DF model \citep[see][]{tremmel:2015aa}. Similarly to \citetalias{Ricarte:2021aa}, within \citetalias{Weller:2022aa_Astrid} the non-central IMBHs are defined as wandering IMBHs.

As mentioned, another method for simulating BHs involves using stellar clusters as proxies: these stellar clusters serve as potential hosts for BHs formed via stellar dynamical processes. By tracing the features and behaviour of these stellar clusters, one can indirectly infer the presence and characteristics of the BHs situated within the clusters. In this work, we applied this approach to the GigaEris simulation, employing comparable selection criteria to those outlined in \citetalias{Weller:2022aa_TNG50}, by selecting stellar clusters acting as BH proxies. \citetalias{Weller:2022aa_TNG50} employs five different sets of IMBHs, each with different conditions. In this study, our focus is specifically on their Sets~2 and 3. For Set~2, a prerequisite is that the IMBH must be within 0.05~ckpc of the dwarf's centre of mass at the snapshot when the dwarf appears in the MW-analog merger tree. Conversely, Set~3 includes all captured IMBHs that have originated outside this MW analog. In both sets, stellar clusters with a stellar mass $10^4$ M${_{\sun}} \leq M_{\star} \leq 10^6$ M${_{\sun}}$ are considered as IMBHs.

It is crucial to note the variations in redshift across the different simulations. The analyses of these simulations are conducted for TNG50 (\citetalias{Weller:2022aa_TNG50}) at $z = 0$, ASTRID (\citetalias{Weller:2022aa_Astrid}) at  $z = 3$, and GigaEris at approximately $z = 4$. The work described by \citetalias{Ricarte:2021aa} has been carried out at different redshifts. For our analysis, we will specifically utilize the findings at both $z \sim 0$\footnote{At this redshift, the BH seeds have an average mass of $\sim$$10^7$~M$_{\sun}$ in a halo of $\sim$$10^{12}$~M$_{\sun}$, which is the selected $z=0$ halo mass of the GigaEris simulation.} and $z = 3$.

\subsubsection{The GigaEris samples}\label{giga_selection}

As previously stated, the criteria used to model IMBHs within the GigaEris simulation are shaped by the methodology outlined in \citetalias{Weller:2022aa_TNG50} by using stellar cluster proxies. To find the stellar clusters within the GigaEris simulation, we employed the adaptive mesh \textsc{AMIGA Halo Finder} \citep[\textsc{AHF};][]{Gill:2004aa, Knollmann:2009aa} to the final snapshot of the simulation at $z = 4.4$. The clusters were selected in an identical way as in \citet{vandonkelaar:2023aa, vandonkelaar:2023ab}, namely with a minimum threshold of 64 baryonic particles within the cluster's virial radius and zero subclusters inside the identified cluster.

Three sets of wandering IMBHs are created. Two of the sets employed slightly modified selection criteria from those outlined in \citetalias{Weller:2022aa_TNG50}, whereas the third set aimed to closely align with the criteria used in simulations employing the BH-seeding method, as illustrated in Table~\ref{tab:GigaEris_sets}. The initial set, Set~A, involves the selection of stellar clusters (functioning as proxies for IMBHs) with a stellar mass\footnote{To be consistent with the analysis performed in \citet{vandonkelaar:2023aa,vandonkelaar:2023ab}, we define the stellar mass of a cluster as the stellar mass within half of its virial radius.} of $10^4$ M${_{\sun}} \leq M_{\star} \leq 10^6$ M${_{\sun}}$, whose birth location lies outside the virial radius of the main galaxy halo, computed at $z = 4.4$, and are captured by the galaxy by that final time. The birth location refers to the centre of mass coordinates where the combined mass of the stellar particles (within a sphere with a maximum radius of 0.2~kpc) recognized as part of the cluster at $z = 4.4$ reaches $10^4$~M$_{\sun}$ for the first time. The 0.2~kpc limit was set to ensure that stellar particles are proximate to each other within the simulation, rather than scattered widely apart. This distance is roughly four times the softening of our simulation and approximately the average virial radius of our clusters. We conducted a trial by altering the `born outside the galaxy' criterion, by considering the virial radius at a couple of varying redshifts. However, this adjustment did not introduce any additional IMBHs to our set. Hence, we consider this to be an acceptable approximation.

\begin{table} 
\centering
\caption{Overview of the selection criteria of the three wandering-BH sets used in this paper.}
\begin{tabular}{c|c|c|c}
\textbf{Set \#} &
  \textbf{Selection Criteria} &
  \textbf{Similar to}  &
  \textbf{BHs} \\ \hline
Set A &
  \begin{tabular}[c]{@{}c@{}}Identified clusters that \\originate outside the MW- \\ analog's virial radius, in the\\ mass range $10^4$--$10^6$~M$_{\sun}$\end{tabular} & Set 3 in \citetalias{Weller:2022aa_TNG50} & 13 \\ \hline
Set B &
  \begin{tabular}[c]{@{}c@{}} Subset of Set A born \\within 0.1 kpc from \\the central region \\of a dwarf galaxy \end{tabular} & Set 2 in \citetalias{Weller:2022aa_TNG50}& 12\\  \hline
Set C &
  \begin{tabular}[c]{@{}c@{}} Non-central clusters\\ within the galaxy with a  \\  stellar mass above $10^6$~M$_{\sun}$ \end{tabular} & \citetalias{Ricarte:2021aa} & 18\\  
\end{tabular}
\label{tab:GigaEris_sets}
\end{table}

This results in 13 captured IMBH proxies, with stellar masses in the range $10^{4.2}$~M${_{\sun}} \lesssim M_{\star} \lesssim 10^{5.8}$~M${_{\sun}}$ and a mean stellar mass of 10$^{5.1}$~M${_{\sun}}$. This approach is similar to that used to obtain Set~3 in \citetalias{Weller:2022aa_TNG50}. However, the GigaEris simulation features a better baryonic mass resolution, and the interpretation of ``born outside the galaxy'' may vary in its definition. Nevertheless, we assume that \citetalias{Weller:2022aa_TNG50} have also adopted the virial radius as their definition for ``outside the galaxy''.

Since most IMBHs are found in the centre of dwarf galaxies in the local Universe \citep[e.g.][]{Kunth:1987aa, Barth:2004aa, Reines:2013aa, Moran:2014aa, Rizzuto:2023aa}, we use the IMBH proxies from the defined Set~A and consider only the stellar clusters which are born within 0.1~kpc away of the centre of mass of a dwarf galaxy, creating Set~B. The 0.1~kpc criterion was chosen to allow for more than one stellar cluster acting as IMBH in the dwarf galaxy. This gave us 12 IMBH proxies, with a mean stellar mass of 10$^{5.0}$~M${_{\sun}}$, telling us that most of the IMBHs in our work are born in the proximity of dwarf galaxies. 

\begin{figure}
\centering
\setlength\tabcolsep{2pt}%
\includegraphics[ trim={0cm 0cm 0cm 0cm}, clip, width=0.45\textwidth, keepaspectratio]{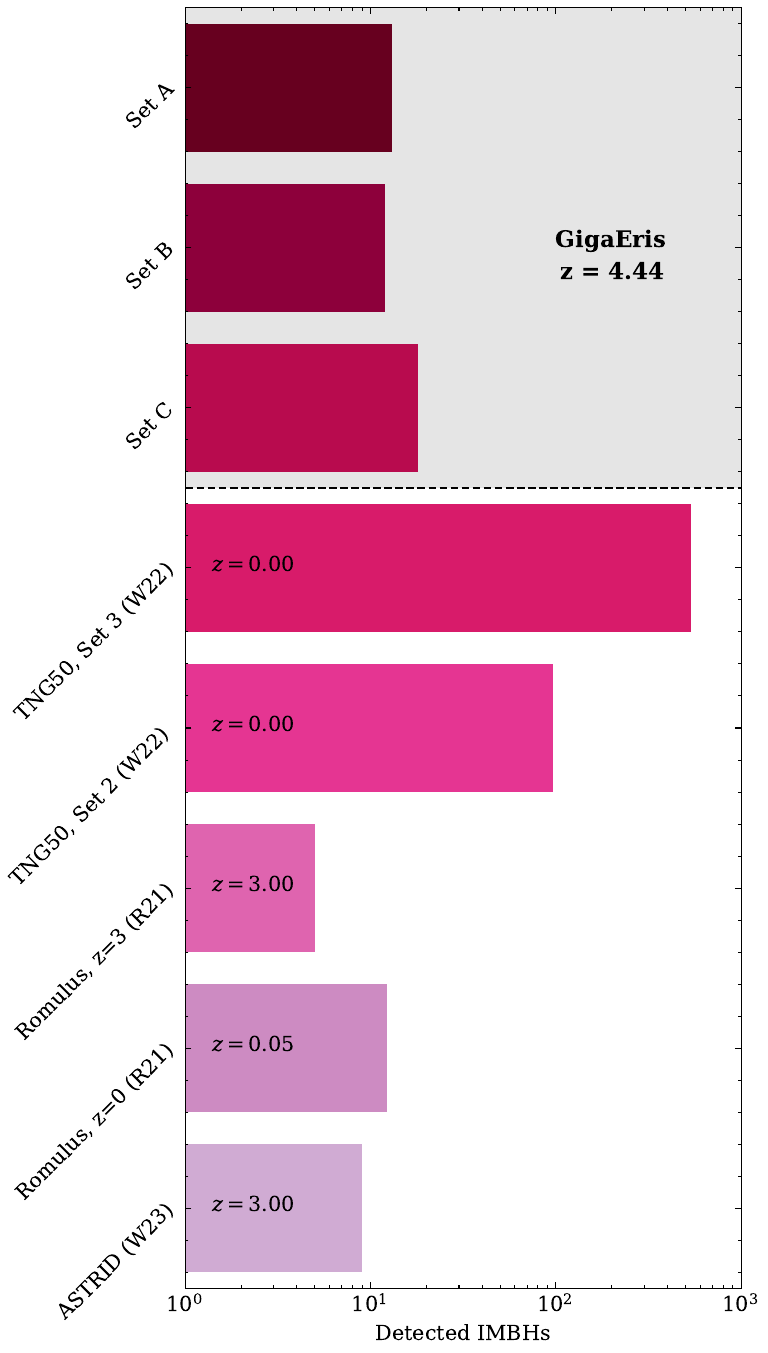}
\caption{The amount of identified wandering IMBHs per MW-mass galaxy in the GigaEris simulation compared with the works discussed in this paper, namely  ASTRID (\citetalias{Weller:2022aa_Astrid}), Romulus (\citetalias{Ricarte:2021aa}), and TNG50 (\citetalias{Weller:2022aa_TNG50}). ASTRID and Romulus both implement BH seeding as simulation technique, characterizing BH dynamics using a consistent sub-grid DF model \citep[see][]{tremmel:2015aa}. In contrast, the TNG50 and GigaEris works utilize stellar clusters to represent BHs. The simulations are analysed at different redshfits, TNG50 (\citetalias{Weller:2022aa_TNG50}) at $z = 0$, ASTRID (\citetalias{Weller:2022aa_Astrid}) at $z = 3$, and GigaEris at $z = 4.44$. The work of \citetalias{Ricarte:2021aa} has been carried out at different redshifts: for our analysis, we use their findings at $z=0.05$ and $z=3$.}
\label{fig:decIMBHs}
\end{figure}

The third set, Set~C, involves the selection of stellar clusters with a stellar mass $ M_{\star} \geq 10^6$~M${_{\sun}}$ with a minimum distance of 0.7~kpc away from the halo centre at $z=4.4$. This approach is similar to that used in \citetalias{Ricarte:2021aa}, as the initial mass of their BH seed is set to $10^6$~M${_{\sun}}$. This set will be used for comparing the different outcomes between BH seeding and stellar cluster proxies as part of a simulation approach. This last set results in 18 captured BH proxies,\footnote{In \citetalias{Ricarte:2021aa}, wandering BHs are defined using a minimum distance of 0.7~kpc, set at twice their gravitational softening length. By applying a similar threshold of twice the softening length here (0.086~kpc in our case), the count of wandering BHs increases from 18 to 20, while the stellar mass range and mean remain unchanged.} with stellar masses in the range $10^{6.0}$~M${_{\sun}} \lesssim M_{\star} \lesssim 10^{8.1}$~M${_{\sun}}$ and a mean stellar mass of 10$^{7.1}$~M${_{\sun}}$. In \citetalias{Ricarte:2021aa}, the average mass of wandering BHs in a halo with a comparable mass at $z=3$ is found to be approximately 10$^{6}$~M${_{\sun}}$, indicating that, by $z=3$, the BH seeds have not significantly accreted additional mass.

As a consequence of these selection criteria, clusters previously examined in \citet{vandonkelaar:2023aa, vandonkelaar:2023ab} can be included within the IMBH sample. Within Set~A and Set~B, three clusters have been incorporated, which were previously recognized as proto-globular clusters in \citet{vandonkelaar:2023aa}. Additionally, Set~C includes five clusters identified as nuclear star cluster predecessors in \citet{vandonkelaar:2023ab}.

\section{Results}\label{sec:results}

To understand the influence of the resolution and the method of identifying/modelling IMBHs within cosmological simulations, we compare the various counts of wandering IMBHs within the works discussed in this study in Figure~\ref{fig:decIMBHs}. The figure illustrates that simulations utilizing the BH-seeding method exhibit a comparable number of detected wandering BHs, albeit of different masses, at a similar redshift. For instance, at a redshift of $z = 3$, \citetalias{Ricarte:2021aa} observed an average of five wandering BHs per galaxy for galaxies within a virial mass bin of around $10^{11}$~M$_{\sun}$. At a similar redshift, \citetalias{Weller:2022aa_Astrid} detected 8, 8, and 12 IMBHs within three MW-like galaxies, resulting in an average of nine IMBHs per MW-like galaxy. Nevertheless, where the BHs within \citetalias{Ricarte:2021aa} have still an approximate mass of 10$^{6}$~M${_{\sun}}$ at $z=3$\, consistent with almost no growth from their initial BH seed mass, the BHs within \citetalias{Weller:2022aa_Astrid} have masses ranging from $\sim$$4.7 \times 10^4$ to $\sim$$2.7 \times 10^6$~M${_{\sun}}$ at the same redshift.

Conversely, there are noticeable disparities between the two simulations using stellar proxies for BHs.\citetalias{Weller:2022aa_TNG50} identified a total of four MW analogs whose centres of mass fall within one of the three TNG50-1 sub-boxes at snapshot 99. Amongst these four analogs, only a single MW-like galaxy\footnote{Galaxy with ID 565089 within TNG50-1 at snapshot 99.} was found to host IMBHs consistent with the criteria of  Set~2 (see Table~\ref{tab:GigaEris_sets}), containing 385 wandering IMBHs. Consequently, on average, 96 wandering IMBHs per MW-like galaxy were found across the four analogs studied, originating no more than 0.05~ckpc away from the centre of mass of a dwarf galaxy. In their Set~3 analysis, which looked for all captured IMBHs that originate outside the MW analog, they concentrated exclusively on this single MW-like galaxy and detected 2148 wandering IMBHs. For consistency with their Set~2 results, we divide this number by 4, obtaining 537 IMBHs per MW-like galaxy. In contrast, within the GigaEris simulation, we find 12 to 18 IMBHs, with no more than 13 wandering IMBHs identified using  selection criteria comparable to \citetalias{Weller:2022aa_TNG50}'s (see Set~A and Set~B in Table~\ref{tab:GigaEris_sets}). The difference in detected results between \citetalias{Weller:2022aa_TNG50} and our work (GigaEris) could have resulted from the differences in particle mass in the two simulations. The particle mass in the GigaEris simulation, detailed in Section~\ref{sec:sim}, is approximately $\sim$$10^3$~M$_{\sun}$, whereas in the TNG50 simulation, it is of the order of $8.5\times 10^{4}$~M$_{\sun}$.\footnote{This mass represents a target baryonic mass, which is approximately equal to the average initial stellar particle mass. It is worth noting that the final baryonic mass might be slightly below the quoted value.} This results in the possibility that a single stellar particle could be a wandering IMBH within the TNG50 simulation, whereas in the GigaEris simulation, a minimum of $\sim$10 stellar particles is required within a stellar proxy to qualify as an IMBH. 

Caution is warranted in interpreting these results, given also the different redshifts between the simulations. Nonetheless, the halo masses, along with the stellar and gas masses of both galaxies, appear to be of similar orders of magnitude. At a redshift of approximately $z\sim 4.4$, the main galaxy halo within GigaEris exhibits a total virial halo mass of $10^{11.88}$~M$_{\sun}$, with $M_{\rm gas} = 10^{10.37}$~M$_{\sun}$ and $M_{\rm star} = 10^{10.43}$~M$_{\sun}$. Conversely, the galaxy analyzed by \citetalias{Weller:2022aa_TNG50} at $z \sim 0$ showcases a total virial halo mass of $10^{11.36}$~M$_{\sun}$, with $M_{\rm gas} = 10^{10.02}$~M$_{\sun}$ and $M_{\rm star} = 10^{10.49}$~M$_{\sun}$. The resemblance between these two galaxies is likely attributed to the early bursty SF within the GigaEris simulation and the deliberate selection of a final halo mass of $\sim$$10^{12}$~M$_{\sun}$. Furthermore, the IMBHs present in the GigaEris simulation are unlikely to reach the centre of their host structures by $z=0$ (see Section~\ref{sec:DF}), suggesting that the counts of 13 and 12 wandering IMBHs in Sets~A and B are lower limits for the number count at $z = 0$. This is because there is a high possibility for more clusters to contribute to this count over time (after $z \sim 4$), for example, due to minor mergers or the infall of stellar clumps (like the proto-globular clusters in \citealt{vandonkelaar:2023aa}).  However, the stark disparity in the IMBH counts between the GigaEris and  \citetalias{Weller:2022aa_TNG50} highlights a significant discrepancy. It is implausible that approximately 100 dwarf galaxies would merge with the main halo in GigaEris simulation to reconcile Set~B's counts with the results observed in Set~2 of \citetalias{Weller:2022aa_TNG50}.

\begin{figure}
\centering
\setlength\tabcolsep{2pt}%
\includegraphics[ trim={0cm 0cm 0cm 0cm}, clip, width=0.44\textwidth, keepaspectratio]{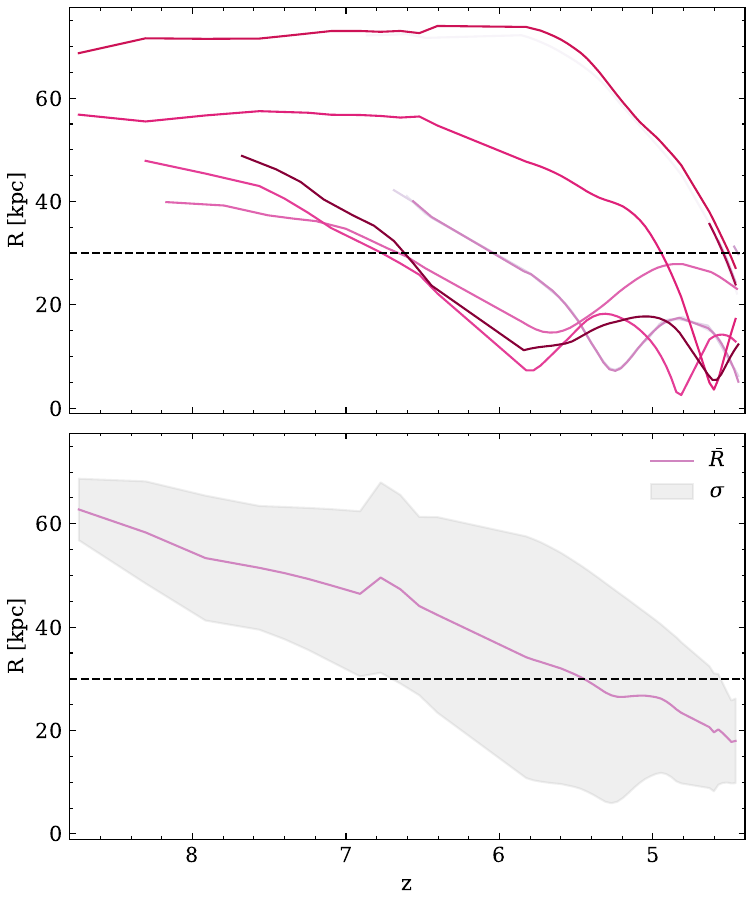}
\caption{Top panel: time evolution in the radial distance from the main galaxy's centre for the 12 IMBHs in Set~B; due to some of the BHs being born in the same dwarf galaxy, some of the lines overlap. The dashed line marks the virial radius at $z=4.4$, used to determine if the IMBH is born outside of the main galaxy. Bottom panel: average radial distance ($\bar{R}$) for the 12 Set~B IMBHs, with the gray shading corresponding to the standard deviation.}
\label{fig:radialT}
\end{figure}

Tantalizingly, we detect a roughly similar amount of wandering IMBHs in the GigaEris simulation using the criteria outlined in Set~C  as the simulations using BH seeding to simulate BHs. Set C is of particular interest, as it uses a similar mass range to the BH seeds used in \citetalias{Ricarte:2021aa}. While Sets~A and B also have similar IMBH counts, they do not represent the same mass range as the BH seeds in \citetalias{Ricarte:2021aa}, making them less directly comparable. This demonstrates a possibility that the utilization of stellar clusters as substitutes for BHs within cosmological simulations yields comparable results to the conventional BH-seeding method, given that the simulation surpasses a distinct mass  resolution threshold. This resolution threshold appears to lie between the resolutions of the TNG50 simulation and the GigaEris simulation. As previously mentioned, the particle mass resolution in the GigaEris simulation is higher compared to the TNG50 simulation.  However, beyond this critical resolution threshold, the detection of IMBHs in systems featuring stellar proxies seems to converge with those incorporating BH seeding. This outcome must be interpreted with caution, since the masses of the BHs amongst the simulations (GigaEris and \citetalias{Ricarte:2021aa}) are different by approximately one order of magnitude (see Section~\ref{sec:criteriamethod}). However, our clusters can also be interpreted as IMBHs surrounded by baryons, hence the difference in mass could be smaller \citep[depending on the model; see Section~\ref{sec:DF} and][]{Fragion:2018aa}.

\subsection{Inward radial evolution}

Since we employ a similar post-processing technique for the selection of IMBHs as described in \citetalias{Weller:2022aa_TNG50}, we can make a comparison between the temporal evolution of radial distances from the centre of the main halo of the IMBHs within the different simulations. Figure~\ref{fig:radialT} indicates that the radial distances of IMBHs in our simulation tend to decrease as time progresses. This result aligns with the conclusions drawn by \citetalias{Weller:2022aa_TNG50}, who concluded that the captured IMBHs exhibit a declining radial distance as time progresses. This effect could be explained by mass migration due to DF. 

\subsubsection{The contribution of intermediate-mass black holes to the supermassive black hole growth}\label{sec:DF}

\begin{figure}
\centering
\setlength\tabcolsep{2pt}%
\includegraphics[ trim={0cm 0cm 0cm 0cm}, clip, width=0.48\textwidth, keepaspectratio]{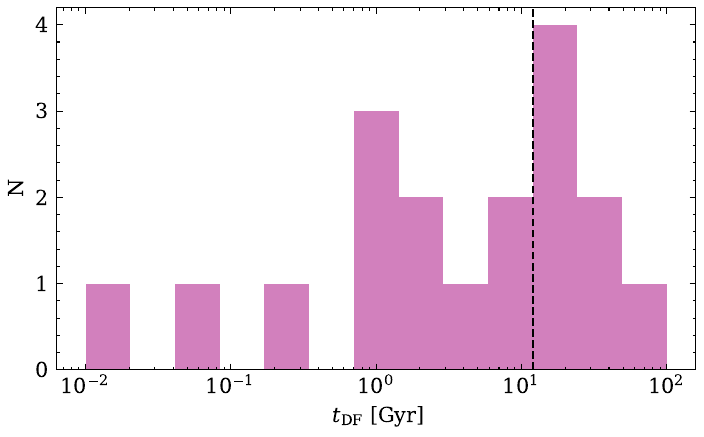}
\caption{The full DF time-scale distribution for the stellar cluster proxies acting as IMBHs within Set~C. The dashed vertical line indicates 12~Gyr, which is the time between the last snapshot in the simulation, at $z = 4.4$, and $z \sim 0$. 11 systems out of 18 have a $t_{\rm DF} < 12$~Gyr.}
\label{fig:tdf}
\end{figure}

The SMBH of the MW is theorized to form through the growth of seed BHs, originating from the collapse of massive gas clouds or the mergers of smaller BHs \citep[e.g.][]{Volonteri:2010aa}. Studies have suggested that IMBHs could undergo a series of mergers and accretion events, gradually evolving into the SMBH observed today \citep[e.g.][]{Alister:2020aa, Abbas:2021aa}.

In an attempt to estimate the time required for the IMBHs in this simulation to migrate to the galaxy's centre, we calculated the DF time $t_{\rm DF}$ based on their positions at $z = 4.4$. We utilize an analytical model akin to that in \citeauthor{vandonkelaar:2023ab} (\citeyear{vandonkelaar:2023ab}; see also \citealt{SouzaLima_et_al_2017,Tamburello_et_al_2017}), by first fitting the inner 30~kpc of the galactic (DM and baryonic) density profile as $\rho_{\rm tot}(r) = \rho_0 (r/r_0)^{-2.5}$, where $\rho_0 = 1.2 \times 10^{9.2}$~M$_{\sun}$~kpc$^{-3}$ and $r_0 = 1$~kpc. The time-scale for an IMBH of mass $M_{\star}$ in motion inside such a matter distribution to decay from an initial radial distance, $r_{\rm i}$, from the galactic centre to a final distance, $r_{\rm f}$, can be estimated using the \citet{chandra:1943aa}'s DF formalism, which yields the drag force acting on the IMBH \citep[see also][]{Binney:2008aa}:

\begin{equation}\label{chandra}
    \mathbf{F}_{\rm DF} = -16 \pi^2 G^2 M^2_{\star} m_{\rm a} \ln{ \Lambda} \left[ \int_0^v  v_{\rm a}^2 f(v_{\rm a}) {\rm d}v_{\rm a} \right] \frac{\mathbf{v}}{v^3},
\end{equation}

\begin{figure*}
\centering
\setlength\tabcolsep{2pt}%
\includegraphics[ trim={0cm 0cm 0cm 0cm}, clip, width=0.99\textwidth, keepaspectratio]{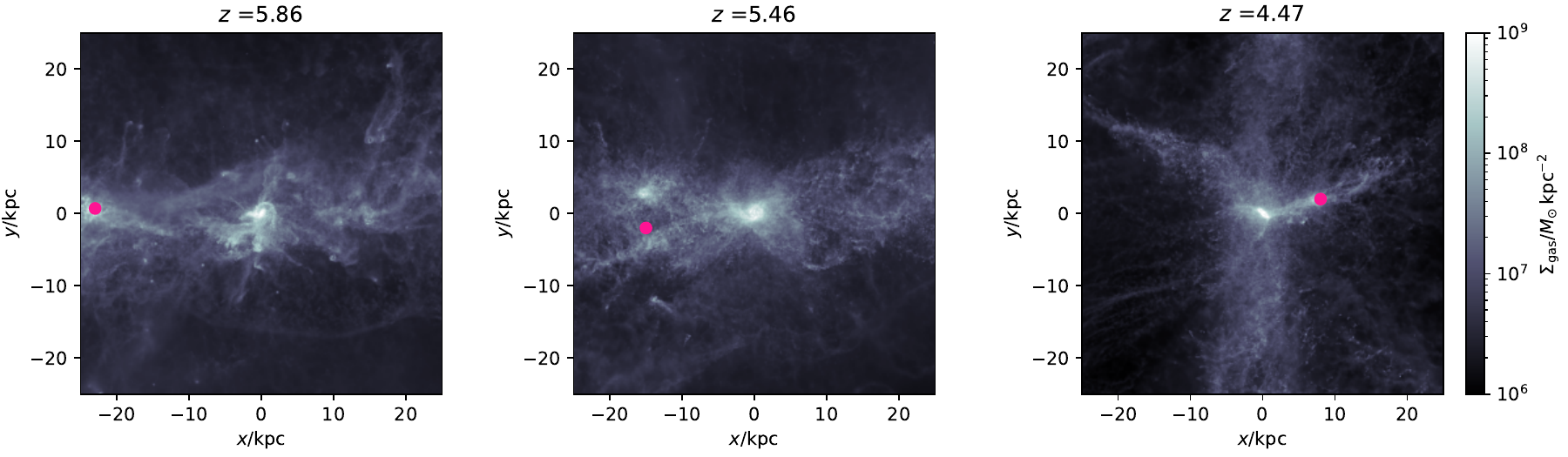}
\caption{Gas surface density maps centred on the main galaxy's halo within the GigaEris simulation, with the birth location and redshift of three wandering IMBHs from Set~C. From the different birth locations, it is clear that these three IMBHs are born within a gas stream towards the centre of the main galaxy's halo.}
\label{fig:birth_Stream}
\end{figure*}

\noindent where $\mathbf{v}$ represents the velocity of the IMBH relative to the background, $m_{\rm a}$ denotes the individual mass of the particles in the background, and $v_{\rm a}$ denotes their velocity. The term $f(v_{\rm a}){\rm d}v_{\rm a}$ signifies the number of particles with velocities between $v_{\rm a}$ and $v_{\rm a}+{\rm d}v_{\rm a}$. Lastly, $\ln\Lambda$ stands for the Coulomb logarithm and $G$ represents the gravitational constant. 

The Coulomb logarithm,  $\ln{\Lambda}$, can be approximated as \citep[][]{Binney:2008aa} $\Lambda \approx (b_{\rm max} v^2_{\rm typ})/(G M_{\rm cluster})$, where $b_{\rm max}$ is the maximum impact parameter, $M_{\rm cluster}$ is the stellar mass computed at half the virial radius \citep[using the same method as described in][]{vandonkelaar:2023aa, vandonkelaar:2023ab} of the identified BH proxy, and $v_{\rm typ}$ is the typical velocity in the system. For each detected IMBH, we take $b_{\rm max} = r_{\rm i}$ and $v_{\rm typ}^2 = G M(<r_{\rm i})/r_{\rm i}$, where $M(<r_{\rm i})$ is the total enclosed mass within $r_{\rm i}$ \citep[][]{Binney:2008aa}, and compute $\Lambda$, thus obtaining the DF time-scale.

Setting $r_{\rm i}$ as the distance at $z = 4.4$ between the IMBH and the galaxy's halo's centre of mass and $r_{\rm f} = 0$, postulating circular motion of the IMBH, and considering an isotropic distribution function for the velocities \citep[all assumed to be smaller than $v$;][]{SouzaLima_et_al_2017}, we find that the IMBHs in Sets~A and B would require between $10^{2.7}$ and $10^{4.8}$~Gyr to decay, suggesting that they are unlikely to reach the central region by $z = 0$. On the contrary, the more massive BHs from Set~C require a significantly shorter time, ranging from $10^{-1.7}$ to $10^{1.9}$ Gyr to reach the centre, as depicted in Figure~\ref{fig:tdf}. This implies that a cumulative stellar mass of $10^{8.3}$~M$_{\sun}$, from 11 stellar cluster proxies acting as BHs with a $t_{\rm DF} < 12$~Gyr (from Set~C), could potentially contribute to the total mass of the SMBH.  

In reality, one could assume that the stellar cluster proxy method effectively represents an IMBH encircled by baryonic mass. This implies that the mass range utilized by \citetalias{Weller:2022aa_TNG50} (and in our Sets~A and B) may be too low, as only a fraction of this mass actually constitutes the IMBH itself. For instance, \citet{Fragion:2018aa} considered that up to 4 per cent of the cluster's mass corresponds to the BH \citep[but see also][]{Portegies:2002aa,Sesana:2012aa, Pestoni:2021aa}. With these numbers, the GigaEris Set~C IMBHs and the \citetalias{Ricarte:2021aa} IMBHs would differ in mass only by a factor of $\sim$2. However, if we use lower estimates \citep[e.g. 0.1 per cent;][]{Portegies:2002aa}, the difference in mass increases, but we remind the reader that the seed mass of \citetalias{Ricarte:2021aa} was large (for IMBHs) to begin with.

Considering the infalling IMBHs from Set~C and using the above mentioned mass fraction limits of 0.1 and 4~per cent, this yields a total contribution to the SMBH between $10^{5.3}$ and  $10^{6.9}$~M$_{\sun}$. The estimated mass of the SMBHs in the MW and M31 \citep[e.g.][]{Pods:2003aa,Farr:2011aa,Elbert:2018aa} fall within a similar mass range, suggesting that the accretion of IMBHs could indeed play a non-negligible role in growing the mass of the SMBH of massive spirals. However, it is important to  note that the mass employed in the DF calculation represents an upper limit under the assumption that only a percentage of the cluster's mass corresponds to the IMBH. This is due to the potential stripping of stars as they migrate towards the centre, resulting in the mass of the cluster being reduced and having a longer the DF time-scale. Consequently, the actual cumulative mass that would contribute to the SMBH would likely be lower.

Nevertheless, further research is needed to determine the importance of wandering IMBHs at $z \sim 4.4$ as seeds for the SMBH. This is especially important in light of prior findings, such as those presented by \citet{Bonoli:2016aa}, who suggested that within galaxies with a mass similar to that of GigaEris, the efficiency of late-time gas accretion (at $z < 6$) seems to be relatively low because, once an extended smooth disc forms and the merger activity decreases, central gas inflows feeding galactic nuclei become very weak. Moreover, our calculations indicate that IMBHs from both Set~A and Set~B, which more closely resemble the mass of an IMBH in isolation, contribute negligibly to the growth of the SMBH.

\subsection{Simulating methods and the formation channels}\label{sec:form}

The methods for modelling IMBHs in simulations are actually associated to two distinct formation channels. One involves the formation of IMBHs through stellar dynamical processes such as runaway collisions amongst stars or other compact objects in dense stellar clusters \citep[e.g.][]{Portiegies:1999, Portiegies:2004aa, Devecchi:2009aa, Mapelli:2016aa, Antonini:2019aa, Rose:2022aa, Purohit:2024aa, Rantala:2024aa}. This formation channel is closely tied to simulating the BHs with the use of stellar cluster proxies. On the other hand, the seeding of BHs focuses on a different formation channel, wherein IMBHs result from the direct collapse of rapidly inflowing dense gas \citep[e.g.][]{Loeb:1994aa, Eisenstein:1995aa, Bromm:2003aa, Lodato:2006aa, Mayer:2023aa}. 

\begin{figure}
\centering
\setlength\tabcolsep{2pt}%
\includegraphics[ trim={0cm 0cm 0cm 0cm}, clip, width=0.45\textwidth, keepaspectratio]{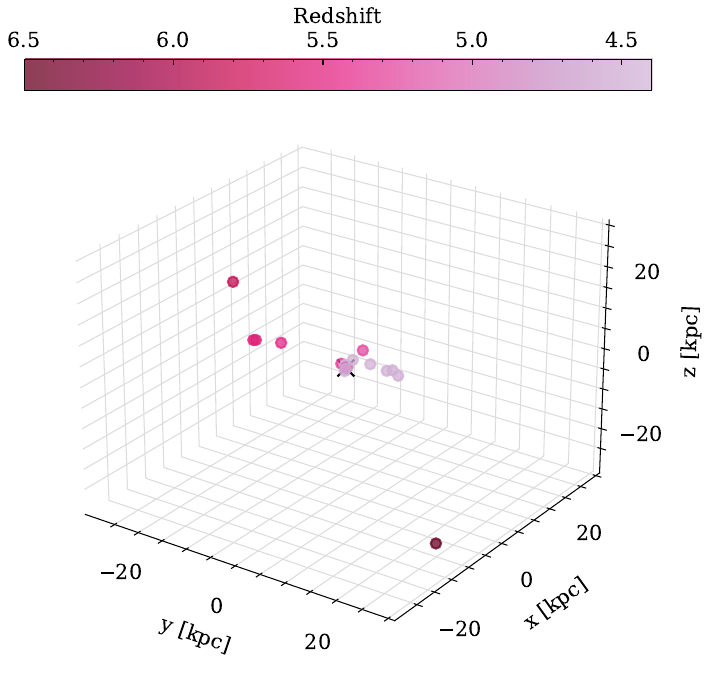}
\caption{The birth location of the IMBHs in Set~C.The colour bar represents the birth redshift of these IMBHs. The positions are normalized, ensuring that the (0,0,0) coordinate always corresponds to the centre of the main halo in the GigaEris simulation.}
\label{fig:locs}
\end{figure}

\subsubsection{Birth Locations}\label{sec:birth}

The contrast in the formation channels underlying the two simulation methods becomes evident when observing the birth locations of the IMBHs in the simulations. Figures~\ref{fig:birth_Stream} and \ref{fig:locs} illustrate the birth locations and redshifts of IMBHs in Set~C. Specifically, Figure~\ref{fig:birth_Stream} displays the birth locations of three randomly selected wandering IMBHs, accompanied by a gas surface density map. The complete set of birth locations accompanied by the gas surface density map for all wandering IMBHs in Set~C is provided in the Appendix, as shown in Figure~\ref{fig:apppendix}. 

From Figure~\ref{fig:locs}, we could infer that most of the wandering IMBHs within the GigaEris simulation typically originate along a single trajectory towards the main galaxy halo. This trajectory coincides with a gas flow directed towards the galaxy, as shown in the right-hand panel of Figure~\ref{fig:birth_Stream}. The relationship between the birth locations of the IMBHs in this work and the regions of gas inflows establishes a significant correlation that links the formation of IMBHs with the initial conditions of stellar cluster formation environments, as found in \citet{vandonkelaar:2023aa}.  They emphasized that the spatial distribution of the detected proto-globular clusters bears a striking resemblance to a filamentary network composed of multiphase gas, a characteristic associated with the cooling process of intracluster gas \citep[see also][]{Lim:2020aa}. The birth location of the wandering IMBHs traces the location of the proto-globular clusters. Moreover, Figure~\ref{fig:locs} demonstrates that most of these wandering IMBHs were formed at redshifts below 5 and are concentrated in a similar region. This region is illustrated in the third panel of Figure~\ref{fig:birth_Stream}. From this panel, one can conclude that these wandering IMBHs originated in or near a dwarf galaxy that is currently falling into the main halo.

Contrary, the stochastic distribution of formation locations observed in the other cosmological simulations using the BH-seeding method, as shown by \citetalias{Ricarte:2021aa} and \citetalias{Weller:2022aa_Astrid}, underscores a noteworthy contrast. The difference is likely the product of the different formation scenario considered in those works and in our paper. We would like to emphasize that both scenarios are equally plausible, but in our scenario IMBHs would form even at late times, as long as dense star clusters are in place in which the various stellar-dynamical channels can be effective at generating IMBHs. The resulting populations of IMBHs would have, by construction, a different evolution with redshift, which however we cannot assess directly due to the limited time-span of GigaEris (but see Section~\ref{sec:DF}). The overall census of IMBHs at low redshift should include both the ``primordial'' population resulting from BH seeds, and a lower-redshift population forming in star clusters.

\section{Discussion and conclusions}\label{sec:disc}

The question of whether current cosmological simulations  can be used to characterize the properties of IMBHs, and guide observational searches for them, is a complex question that depends on the interplay of various factors. Stellar proxies, while serving as a promising option for placing IMBHs within cosmological simulations when considering the number of detections,  yield results that depend strongly on resolution. While we can by no means claim convergence even at the exquisite resolution of GigaEris, the trend is such that the number of IMBHs using stellar cluster proxies as identification methods decreases significantly with increasing resolution, suggesting that using this method in cosmological volume simulations, which by construction have low mass resolution, is amenable to a large inaccuracy.

This issue becomes even more evident when applying the same analysis to the original Eris simulation \citep{Guedes:2011aa}. Eris and TNG50 have comparable resolution, with Eris (TNG50) having an initial stellar particle mass of $6.2 \times 10^3$~M$_{\sun}$ ($8.5 \times 10^4$~M$_{\sun}$) and a softening length of 120~pc (288~pc). Despite this, their outcomes are strikingly different: in Eris we identified zero star clusters (using similar criteria\footnote{Due to the lower resolution, we had to adjust the minimum threshold for baryonic particles within the cluster's virial radius from 64 to 2.} to those of Set~A for GigaEris) both at $z = 0$ and $z = 4.4$, whereas TNG50 revealed 385 clusters (using the selection criteria of Set~3 in \citetalias{Weller:2022aa_TNG50}; see Table~\ref{tab:GigaEris_sets}) at $z = 0$. This might just mean that the halo finder, used successfully in GigaEris, is not able to find halos with this low resolution. Another test was to check how many  star particles (of mass $M_{\star} \approx 6 \times 10^3$~M$_{\sun}$) reside within the virial radius but originated outside the main halo in the Eris simulation. Under these new criteria, we identified 401 star particles at $z=4.4$ and 1226 star particles at $z=0$. The $z=0$ count is of the same order of magnitude as that of the IMBHs identified in the single MW-like galaxy of the TNG50 simulation (Set~3) in \citetalias{Weller:2022aa_TNG50}. Therefore, we argue that this approach -- using a stellar proxy for BHs based solely on the criteria of formation mass and location outside the virial radius -- can be acceptable only at high enough resolution. Nevertheless, using a single stellar particle representation of an IMBH is flawed and should be avoided: there is no guarantee that these single particles genuinely correspond to stellar clusters rather than simply a group of field stars. This contrasts with the stellar clusters identified in our GigaEris simulation, which were explicitly detected as bound clusters through AHF. Assuming that a single stellar particle represents a cluster risks oversimplifying the intricate processes involved, which could result in an overestimation of wandering IMBHs.

The order-of-magnitude agreement in the number of detected wandering IMBHs across a wide range of simulations using different sub-grid methods to identify IMBHs might seem encouraging, but it is only coincidental. Indeed, a significant contrast becomes apparent when examining the formation sites of the wandering IMBHs. The stark difference between the stochastic formation locations seen in the simulations using BH seeding \citep[see \citetalias{Ricarte:2021aa} and \citetalias{Weller:2022aa_Astrid}, but also][]{Bellovary:2010aa, Tremmel:2018aa, Tremmel:2018ab, Pfister:2019aa, Bellovary:2021aa, Chen:2022aa} and those using star clusters as proxies reflects the different underlying formation channel adopted in the two methods. This difference highlights the importance of recognizing that the identified wandering IMBHs using different sub-grid modelling methods correspond to different populations of IMBHs, which in principle are all plausible, but could possibly be told apart in observational searches because they lead to different properties, the spatial distribution being just one of them, and one very accessible from the analysis of cosmological simulations. For instance, the IMBHs in Set~B essentially represent the central BH of a dwarf galaxy that merges with the main galaxy halo, whereas those in Set~C represent a globular cluster with an IMBH inside. Additionally, such different properties and dynamical histories underscore the need for caution, as we may be comparing entirely different objects amongst the different simulations. 

In summary, it is important to exercise extreme caution when interpreting the census, kinematics, and dynamics of wandering IMBHs in the current cosmological simulations. As simulations advance in accuracy and computational power, there is a pressing need to improve, in parallel, our understanding of the sub-grid methods used for representing IMBHs. The stellar-proxy method, guided by criteria outlined in \citetalias{Weller:2022aa_TNG50}, demonstrates promise when paired with sufficiently high resolution. Unlike the BH-seeding method, which would often leave IMBHs isolated, the stellar cluster proxy approach implies that IMBHs would be surrounded by baryonic matter, such as stars, because this reflects the underlying formation mechanism. Although in reality the specific mechanisms can vary, they all rely on collisional dynamical processes inside star clusters. IMBHs forming this way would generally be concealed within clusters, unless the latter has been completely tidally stripped, such as perhaps it is the case for $\omega$ Cen \citep[see, e.g.][]{Noyola:2010aa}. In general, this makes it challenging to identify them, as they reside in ``dry'' hosts, hence would undergo little or no accretion. In the latter case, only dynamical methods can reveal their existence, which poses notorious challenges in absence of accurate 3D information on stellar kinematics, and would explain the lack of solid detections of IMBHs in the local Universe. We argue that, if paired with sufficiently high resolution in the baryonic component, the stellar-proxy method appears to be the most promising approach for guiding searches of IMBHs in clusters as the results on the spatial distribution, numbers, and masses of IMBHs could be used to inform observational searches using high-resolution kinematics in local galaxies. However, a large uncertainty lies in what fraction of mass to assign to the cluster and what fraction of mass to the IMBH. In fact, assuming that 1~per cent of the stellar mass of a stellar proxy represents an IMBH, \citetalias{Weller:2022aa_TNG50} should have identified star clusters with masses in the range of $10^6$ to $10^8$~M${_{\sun}}$ as wandering IMBHs. Coincidentally, this mass range closely matches the one used in our Set~C.

In conclusion, whether or not current cosmological simulations can be used to identify and study wandering IMBHs, and derive their properties, is at the moment a highly uncertain matter. While simulations provide valuable general insights, their role in aiding future  observational searches in the local Universe, including in our Galaxy, requires a careful assessment of the resolution requirements and of the sub-grid methods, with the two being connected. To the very least, simulations should take into account both the BH seeding formation channel and stellar cluster formation channel, and characterize both populations as a function of redshift, as both are possible scenarios which, as we have shown, predict different population properties despite the similar predicted census. Furthermore, in the future, within the stellar cluster proxy method the different dynamical mechanisms to form IMBHs inside clusters should eventually be considered at the sub-grid level, provided that enough information is available in the simulations to differentiate between them, which is of course related to the available resolution. Additionally, one needs to establish a clear and consistent definition of ``wandering IMBHs'' in the literature to ensure effective communication and comparability between future studies. We suggest that the definition of ``non-central BH'', as used by \citetalias{Ricarte:2021aa} and \citetalias{Weller:2022aa_Astrid}, is the most useful, as it provides a practical framework that can be directly compared with observations.

The analysis presented in this paper is by no means exhaustive but should serve as a reference to highlight the challenges involved in this type of research. With the prospect of LISA and the Einstein Telescope probing, in the next decade, both the high- and the low-mass end of IMBH mergers in the GW domain, developing robust population models for IMBHs will be increasingly important. Indeed, wandering IMBHs and  IMBH mergers are tightly connected because wandering IMBHs will be the outcome when IMBH pairing via dynamical processes is inefficient, preventing IMBH mergers. Advancing the modelling in this area is thus relevant for both electromagnetic and GW observations.

\section*{Acknowledgements}
PRC acknowledges support from the Swiss National Science Foundation under the Sinergia Grant CRSII5\_213497 (GW-Learn). LM and FvD acknowledge support from the Swiss National Science Foundation under the Grant 200020\_207406. The simulations were performed on the Piz Daint supercomputer of the Swiss National Supercomputing Centre (CSCS) under the project id s1014. We acknowledge Michael Tremmel for insightful discussions. 

\section*{Data Availability}
The data underlying this article will be shared on reasonable request to the corresponding author.



\bibliographystyle{mnras}
\bibliography{example} 

\appendix
\section{Set~C Birth Locations}
\begin{figure*}
\centering
\setlength\tabcolsep{2pt}%
\includegraphics[ trim={0cm 0cm 0cm 0cm}, clip, width=0.78\textwidth, keepaspectratio]{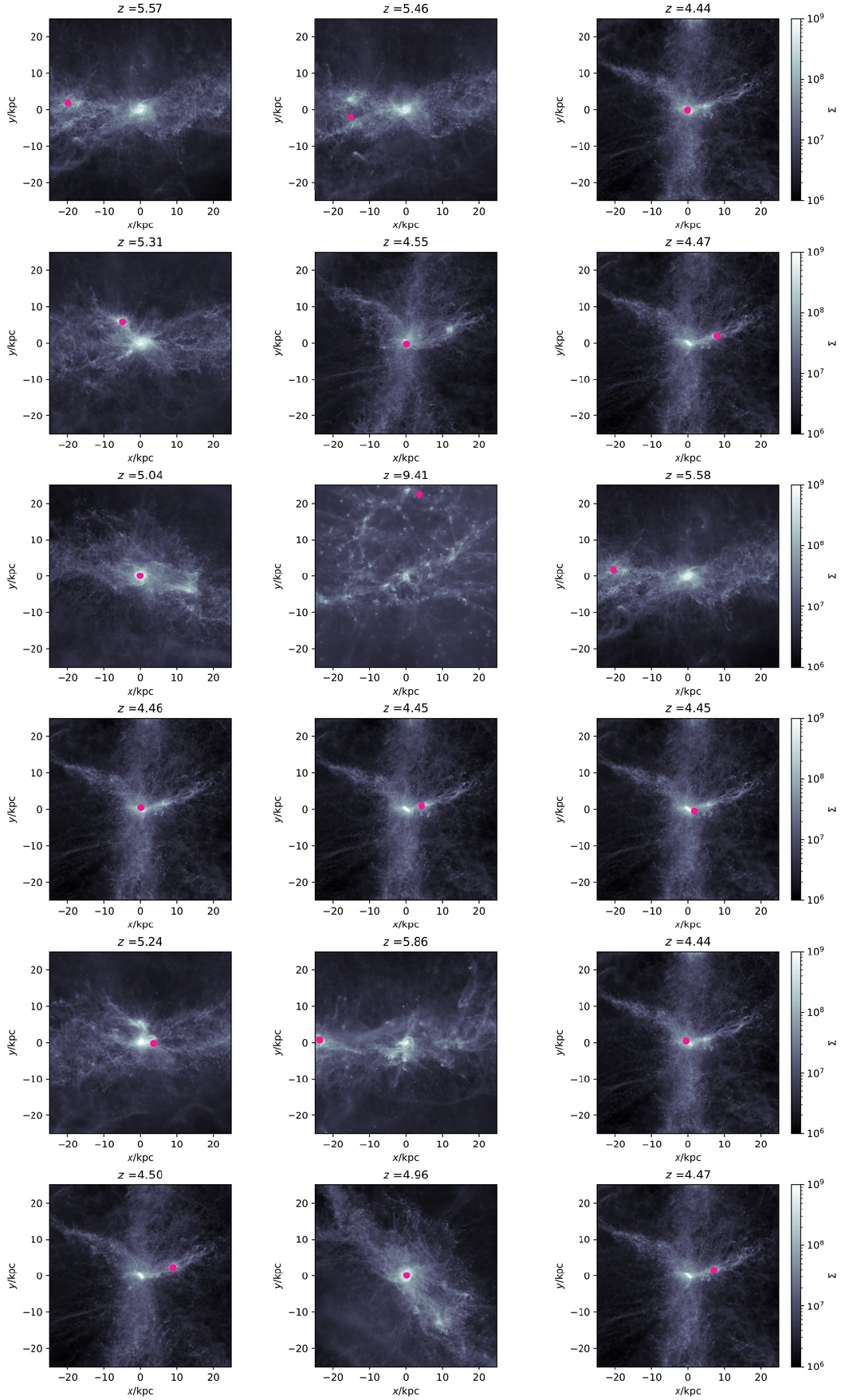}
\caption{Gas surface density maps centred on the main galaxy's halo within the GigaEris simulation, with the birth location and redshift of all the wandering IMBHs from Set~C.}
\label{fig:apppendix}
\end{figure*}

\bsp	
\label{lastpage}
\end{document}